\documentclass[apj,twocolumn,twocolappendix,appendixfloats]{openjournal}
\usepackage{lipsum}
\usepackage{newtxtext,newtxmath,enumitem,multirow}
\usepackage{tabularx,booktabs}
\usepackage{xcolor}
\usepackage{textgreek}
\usepackage[utf8]{inputenc}
\usepackage[english]{babel}
\usepackage{hyperref}
\hypersetup{
    colorlinks=true,
    linkcolor=blue,
    filecolor=blue,      
    urlcolor=blue,
    citecolor=blue,
}
\usepackage{color,colortbl}
\usepackage{tensind}
\tensordelimiter{?}
\DeclareGraphicsExtensions{.bmp,.png,.jpg,.pdf}
\usepackage{verbatim}
\usepackage[normalem]{ulem}
\usepackage{orcidlink}
\usepackage{soul}
\usepackage{bm}
\usepackage{graphicx}
\usepackage{threeparttable}
\usepackage{adjustbox,lipsum}
\usepackage{caption}
\usepackage{comment}
\usepackage{amsmath}
\usepackage{tgbonum}
\usepackage[T1]{fontenc}

\def\ion#1#2{#1$\;${\small\rm\@{#2}}\relax}

\newcommand{\cmjj}{\mbox{${\rm cm^{-2}}$}}

\newcommand{\kms}{\mbox{km\ s${^{-1}}$}}

\graphicspath{ {./figs/} }

\begin{document}
\title{Non-equilibrium ionization in the multiphase circumgalactic medium---impact on quasar absorption-line analyses}

\author{Suyash Kumar$^{1}$\orcidlink{0000-0003-4427-4831}}
\author{Hsiao-Wen Chen$^{1,2}$\orcidlink{0000-0001-8813-4182}}
\thanks{Corresponding author: Suyash Kumar} \email{suyashk@uchicago.edu}
\affiliation{$^{1}$Department of Astronomy and Astrophysics, The University of Chicago, Chicago, IL 60637, USA}
\affiliation{$^{2}$Kavli Institute for Cosmological Physics, The University of Chicago, Chicago, IL 60637, USA}

\begin{abstract}
This paper presents an updated framework for studying the ionizing conditions and elemental abundances of photoionized, metal-enriched quasar absorption systems. The standard assumption of ionization equilibrium invoked in absorption line analyses requires gas to cool on longer timescales than ionic recombination ($t_\mathrm{cool} \gg t_\mathrm{rec}$). However, this assumption may not be valid at high metallicities due to enhanced cooling losses. This work presents a suite of time-dependent photoionization (TDP) models that self-consistently solve for the ionization state of rapidly cooling gas irradiated by the extragalactic ultraviolet background (UVB). The updated framework explores various revised UVBs from recent studies, a range of initial temperatures, and different elemental abundance patterns to quantify the effects of TDP on the observed ion fractions. A metal-enriched ($\mathrm{[\alpha/H]}=0.6_{-0.1}^{+0.2}$) \ion{C}{IV} absorption system at $z \sim 1$ previously studied using photoionization equilibrium (PIE) models is re-examined under the TDP framework. The main findings are as follows: (1) varying prescriptions for the underlying UVB or adopting initial temperatures $T_0 \lesssim 10^6 \ \mathrm{K}$ (with the starting ionization state in collisional equilibrium) change TDP ion fractions by up to a factor of three and ten respectively, but the adopted relative elemental abundance pattern affects ion fractions by at most 40\%; (2) the inferred gas densities are consistent between PIE and TDP, but under TDP solar metallicity cannot be ruled out at more than 2-$\sigma$ significance and a non-solar [C/$\alpha$]$\approx 0.25$ is robustly constrained from the observed relative ion abundances. Extending the TDP analyses to a larger sample of super-solar absorption components with high signal-to-noise absorption spectra is needed to quantify the fraction of metal absorbers originating in rapid cooling gas.

\end{abstract}


\maketitle

\section{Introduction} 

Quasar absorption line spectroscopy provides a powerful tool for probing tenuous gas reservoirs in galaxies. Studies using this technique have led to detections of gaseous clouds out to virial radii and beyond \citep[see e.g.,][]{Chen:2001,Chen:2010,Borthakur:2013,Liang:2014,Huang:2016, Huang:2021, Burchett:2016,Burchett:2019}, which constitute the circumgalactic medium (CGM). Absorption systems have been found to host low-, intermediate-, and high-ionization species, sometimes simultaneously \citep[see e.g.,][]{Chen:2000,Meiring:2013,Burchett:2015,Nevalainen:2017,Zahedy:2019, Zahedy:2021,Sankar:2020,Haislmaier:2021,Cooper:2021,Qu:2023,Qu:2024,Sameer:2024, Kumar:2024}. This indicates a vast range of thermodynamic properties in the CGM; specifically, the CGM exists in the cool ($T\approx10^{4-4.5} \ \mathrm{K}$), warm-hot ($T\approx10^{4.5-5.5}\ \mathrm{K}$), and hot phase ($T\gtrsim10^{5.5} \ \mathrm{K}$). These phases also differ in their baryon and metal budget \citep[see e.g.,][]{Werk:2014,Peeples:2014}, besides their thermodynamic properties and spatial extent.

Thermodynamic properties and elemental abundances cannot directly be inferred for absorption arising from ionized gas. Therefore, ionization models are necessary to infer gas properties from absorption line measurements. To do so, measured relative abundances of ionic species are compared with expectations from ionization models across a grid of gas properties. Available choices of ionization models vary in their incorporation of ionizing radiation. For instance, because collisional interactions are subdued at lower temperatures, the cool and warm-hot CGM are considered to be photoionized by the extragalactic ultraviolet background \citep[see e.g.,][]{Bergeron:1986, Rauch:1997, Keeney:2013, Keeney:2017, Werk:2014,Crighton:2015, Zahedy:2019, Sankar:2020, Zahedy:2019, Zahedy:2021, Cooper:2021,Haislmaier:2021, Qu:2022, Qu:2023, Sameer:2024}. In contrast, hot gas is assumed to be collisionally ionized because of having a higher temperature \citep[see e.g.,][]{Tripp:2011, Meiring:2013, Hussain:2015, Pachat:2017, Nevalainen:2017, Rosenwasser:2018}. 

Both photo- and collisional ionization models typically assume equilibrium conditions. Under this assumption, the gas must cool slowly relative to ionic recombination timescales, so a steady state can be obtained. However, at high metallicities, cooling losses are enhanced through collisional, free-bound, and free-free interactions; the cooling time becomes shorter than ionic recombination, resulting in a departure from ionization equilibrium \citep[see e.g.,][]{Gnat:2007}. In such cases, a steady state cannot be assumed, and equations of ionization balance must be solved simultaneously with energy loss to compute time-dependent ion fractions.

Studies of hot gas have considered both collisional equilibrium and non-equilibrium models \citep[see e.g.,][]{Tripp:2011,Meiring:2013,Hussain:2015,Pachat:2017,Nevalainen:2017,Rosenwasser:2018}. The cool phase has been established to be in photoionization equilibrium \citep[see e.g.,][]{Zahedy:2019,Qu:2022,Sameer:2021,Zahedy:2021,Cooper:2021}, justified by the inferred sub-solar metallicities which lead to slow net cooling. Photoionization equilibrium (PIE) models have also been used for the warm-hot phase \citep[see e.g.,][]{Sankar:2020, Kumar:2024}, but the inferred metallicities at a solar level or above pose a challenge because of enhanced cooling losses expected at these enrichment levels. Additionally, \cite{Kumar:2024} find a discrepancy between the ionization model temperature and the observed thermal broadening in \ion{C}{IV} absorbers after accounting for non-thermal motions in the gas. These caveats motivate the usage of non-equilibrium photoionization models for metal-enriched warm-hot gas.

Non-equilibrium (or time-dependent) photoionization by the extragalactic ultraviolet background (UVB) has been discussed by \cite{Oppenheimer:2013} and \cite{Gnat:2017}. The observed ionization state of a gas out of equilibrium can depart significantly from a gas in equilibrium, assuming otherwise identical physical properties (density, temperature, and elemental abundances). Therefore, systematic differences in inferred gas properties may exist using photoionization equilibrium versus non-equilibrium models. However, before comparing non-equilibrium models with absorption line measurements, some systematics regarding this class of models are worth highlighting.

The primary ingredient for photoionization models is the assumed ionizing radiation background. The CGM, being spatially extended, remains largely unaffected by ionizing radiation arising locally within the galaxy halo \citep[e.g.,][]{Qu:2023}. Instead, the CGM is ionized by the UVB, which has contributions from quasars and star-forming galaxies. There are various prescriptions available for the ionizing background \citep[see e.g.,][]{Haardt:2001,Haardt:2012,Khaire:2019,F-G:2020}; these prescriptions differ in their amplitude and slope, influencing ion fractions in photoionized gas \citep[see e.g.,][]{Chen:2017, Zahedy:2019, Lehner:2022}. Non-equilibrium models from \cite{Oppenheimer:2013} and \cite{Gnat:2017} use UVB prescriptions from \cite{Haardt:2001} and \citet[][HM12]{Haardt:2012}. In the past decade, extensive efforts have been directed towards revising UVB models to incorporate updated statistics of Ly$\alpha$ forest absorbers and improved constraints on AGN luminosity functions/x-ray extragalactic background. As such, recent prescriptions from \cite[][KS19]{Khaire:2019} and \cite[][FG20]{F-G:2020} have not been incorporated into time-dependent photoionization (TDP) models. In addition to the ionizing background, the adopted initial temperature from which the gas is assumed to cool is another important ingredient in non-equilibrium ionization models. This is because the initial temperature sets the starting ionic composition of the gas, which affects the subsequent ionization state of the gas through time-dependent cooling. Finally, the adopted elemental abundance pattern sets the relative cooling losses from individual elements \citep[see e.g.,][]{Gnat:2012}, which are coupled to the ion fractions. It is crucial to explore the variance in non-equilibrium ion fractions resulting from uncertainties in these three model assumptions.

The additional dependence of ion fractions on gas temperature and metallicity under non-equilibrium photoionization is important to address when comparing model predictions with absorption line measurements. In absorption line systems with species of sufficiently different mass and similar ionization stage, the line broadening can be decomposed into thermal and non-thermal contributions \citep[see e.g.,][]{Rauch:1996CIV,Rudie:2019,Zahedy:2019, Qu:2022, Kumar:2024}. In situations where the gas temperature can be determined using the observed absorption line widths, the assumed temperature in the photoionization calculations can be fixed to the line width temperature to reduce degeneracies between different gas properties during ionization modeling \citep[see e.g.,][]{Sameer:2024}.

This work presents a new suite of time-dependent photoionization models with flexible choices for the adopted radiation background, initial temperature, and elemental abundance pattern. The impact of varying model assumptions on non-equilibrium fractions is explored. Following this, a grid of time-dependent photoionization models is compared with absorption-line measurements of the highest metallicity \ion{C}{IV} absorber from \cite{Kumar:2024}. Finally, gas properties (density, metallicity, and relative abundances) inferred using non-equilibrium models are compared with those from equilibrium models.
 
This paper is organized as follows. In \S\ \ref{sec:ion_models}, the construction of ionization models are presented, alongside the impact of varying individual model assumptions on non-equilibrium fractions. In \S\ \ref{sec:analysis}, model predictions are compared with absorption line measurements. In \S\ \ref{sec:discussion}, a summary of the key findings from the study is presented alongside caveats.

\section{Model analysis} 
\label{sec:ion_models}

Recall that in \cite{Kumar:2024}, the authors analyzed a sample of \ion{C}{IV}-selected absorbers identified at $z\sim1$ in high-resolution QSO spectra and derived constraints for the density of the absorbing gas based on a suite of ionization transitions.  The high spectral resolving power enabled the authors to decompose the observed line widths into thermal and non-thermal components, placing strong constraints on the gas temperature in the range of $T \approx 1$--$5\times 10^4 \ \mathrm{K}$ and turbulent velocities of $b_\mathrm{NT} \lesssim 30 \ \kms$. By comparing observed relative ionic abundances of kinematically matched absorbing components with the expectations from photoionization equilibrium models, they showed that \ion{C}{IV} absorption primarily traces diffuse gas of density $n_\mathrm{H} \approx 10^{-4}$--$10^{-3} \ \mathrm{cm}^{-3}$ that is enriched with metallicity of $\mathrm{[\alpha/H]} \gtrsim -1$ (with oxygen, silicon, and neon as tracers of $\alpha$ elements) but with an enhanced carbon abundance of $\mathrm{[C/\alpha]} \gtrsim 0$.  However, this analysis also uncovered two \ion{C}{IV} components for which super solar metallicities ($\mathrm{[\alpha/H]}>0.4$) and a cooler gas temperature than inferred from the observed line profiles are found \citep[e.g., component 3 of the absorber at $z_\mathrm{abs} = 1.26$ in][]{Kumar:2024}.  To explore the possibility of the gas being out of equilibrium due to rapid cooling, here we examine the distinct features associated with equilibrium and non-equilibrium photoionization models and discuss their associated caveats.

\begin{figure} 
    \centering    
    \includegraphics[width=0.9\columnwidth]{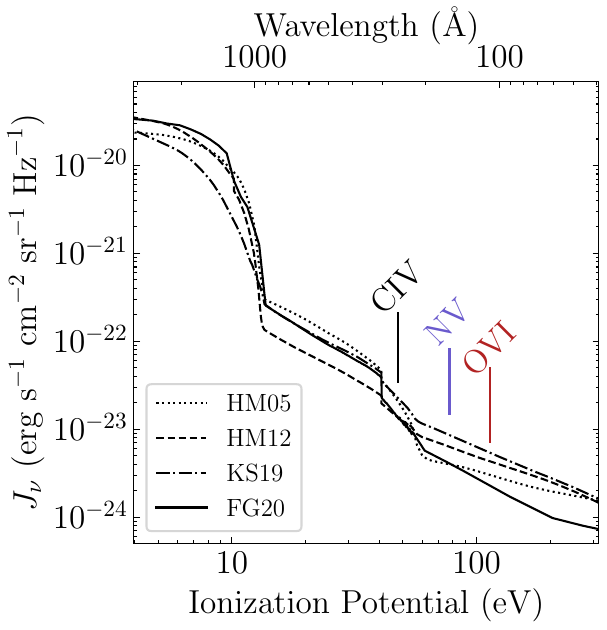}
  \caption{Differences in the expected UVB at $z=1$ based on commonly adopted prescriptions. HM05 refers to an updated version from \cite{Haardt:2001} made available in \texttt{CLOUDY}, while HM12, KS19, and FG20 refer to \cite{Haardt:2012}, \cite{Khaire:2019}, and \cite{F-G:2020}, respectively. The ionization potentials for \ion{C}{IV}, \ion{N}{V}, and \ion{O}{VI}---key tracers of warm-hot gas --- are also marked.} \label{fig:UVB_compare}
\end{figure}

\subsection{Photoionization equilibrium (PIE)} \label{sec:PIE}
 
The photoionization models presented in this paper are calculated using an updated \texttt{CLOUDY} photoionization code \citep[version 22.01, see][]{Ferland:2017,Chatzikos:2023} for a plane-parallel slab of gas irradiated by the extragalactic UVB. Some available UVB prescriptions are shown in Figure \ref{fig:UVB_compare}. The ionization state of gas in PIE is set by a balance between photoionization and ionic recombination.  The equilibrium calculations are performed over a grid of temperatures while including ionizing radiation from the extragalactic UVB, evaluating the ionization state and gas cooling at each temperature independently.

\begin{figure} 
    \centering
    \includegraphics[width=0.9\columnwidth]{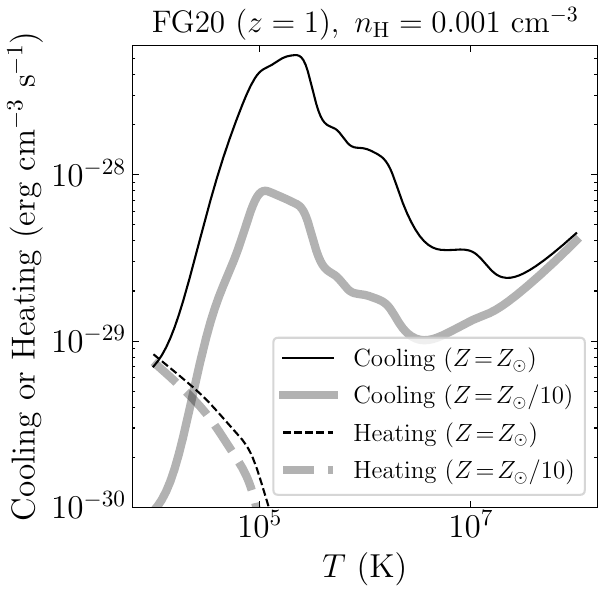}
  \caption{The expected cooling and heating under photoionization equilibrium \citep[see also][]{Wiersma:2009}. A gas density of $n_\mathrm{H} = 0.001 \ \mathrm{cm}^{-3}$ is assumed and the UV background from \cite{F-G:2020} is adopted. Models for solar and one-tenth-solar gas metallicities are depicted using thin and thick lines, respectively.} \label{fig:heat_cool}
\end{figure}

When comparing model predictions with absorption line measurements, the gas is typically assumed to be in thermal equilibrium. Figure \ref{fig:heat_cool} depicts the cooling and heating rates for gas at solar and sub-solar metallicities, assuming a fiducial gas density of $n_\mathrm{H} = 0.001 \ \mathrm{cm}^{-3}$. Note that the metallicity $Z$ for the model curves in Figure \ref{fig:heat_cool} refers to the total gas-phase metallicity, assuming a solar abundance pattern between elements. The equilibrium temperature depends on the gas metallicity, with higher enrichment levels leading to lower temperatures. The equilibrium temperature corresponding to the best-fit ionization model can be compared with the temperature estimated by comparing the absorption line widths from ionic species with sufficiently different masses.

\begin{figure*} 
  \centering
  \includegraphics[width=0.8\textwidth]{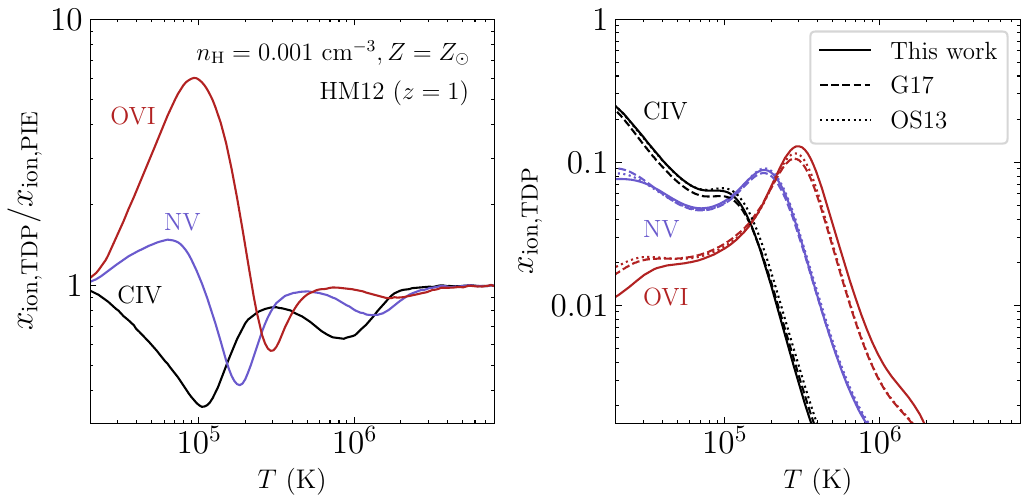}
  \caption{Impact of time-dependent photoionization in the predicted ion fractions based on the framework presented in this paper and those from previous studies. The \textit{left} panel shows differences in the predicted ion fractions $x_{\rm ion}$ between TDP and PIE models for \ion{C}{IV}, \ion{N}{V}, and \ion{O}{VI}. Different species are indicated with different colors. Gas density and metallicity are kept fixed at $n_\mathrm{H} = 0.001 \ \mathrm{cm}^{-3}$ and $Z=Z_\odot$ respectively. The ionizing radiation is assumed to be from 
  HM12 to facilitate comparisons with literature. The \textit{right} panel contrasts the predicted non-equilibrium $x_{\rm ion}$ from this work with those from \cite{Gnat:2017} and \cite{Oppenheimer:2013}. The expected TDP $x_{\rm ion}$ are largely consistent between these studies with differences attributed to updated atomic data.} \label{fig:TDP_PIE}
\end{figure*}

\subsection{Time-dependent photoionization (TDP)} \label{sec:TDP}

In equilibrium conditions, ion fractions ($x_\mathrm{ion}$) are assumed to be in a steady state at all times ($\mathrm{d}x_\mathrm{ion}/\mathrm{d}t\!=\!0$) because the gas cools slowly. However, chemically enriched gas clouds are expected to experience rapid cooling through collisional, free-bound, and free-free interactions.  Therefore, the gas becomes "overionized" relative to the equilibrium scenario and a steady state no longer holds. In such cases, equations of ionization balance must be solved simultaneously with energy conservation to specify $x_\mathrm{ion}$ over time. 

Non-equilibrium collisional ionization models from \cite{Gnat:2007} have been compared with absorption line systems arising from hot gas \citep[see e.g.,][]{Tripp:2011,Meiring:2013,Hussain:2015,Nevalainen:2017,Rosenwasser:2018}. However, photoionization from the extragalactic UVB cannot be neglected at temperatures relevant for some high-ionization species. The inclusion of a radiation background in time-dependent calculations modifies the ionization state of gas through photoionization and photoheating \citep[see e.g.,][]{Oppenheimer:2013,Gnat:2017}. 

TDP calculations for this work are performed in \texttt{CLOUDY} following the \texttt{hazy1} documentation, which provides a non-equilibrium cooling calculation script for modeling the radiative time-dependent cooling in a plane parallel gas slab without any ionizing radiation \citep[see e.g.,][]{Gnat:2007}. The default script is modified to include ionizing radiation from a UVB \footnote{See \href{https://github.com/suyashk12/cgm_science/blob/main/TDP\_utilities/TDP\_readme.txt}{this ReadMe} for a tutorial on building a suite of TDP models.}. Cooling is assumed to be isochoric, and the calculations stop at thermal equilibrium or $10^4 \ \mathrm{K}$, whichever condition is achieved first. The gas is also assumed to be optically thin for these calculations.

\subsection{Comparisons between PIE and TDP} \label{sec:PIE-TDP}

A comparison between PIE (\S\ \ref{sec:PIE}) and TDP ion fractions is illustrated in the left panel of Figure \ref{fig:TDP_PIE} for three commonly adopted tracers of warm-hot ionized gas, \ion{C}{IV}, \ion{N}{V}, and \ion{O}{VI}, in QSO absorption-line studies. Note that the PIE fractions vary with gas density but do not depend on metallicity, while the predicted TDP fractions have a simultaneous density and metallicity dependence. The discrepancy between TDP and PIE is therefore metallicity dependent. At high temperatures, the metallicity-dependence is weaker under TDP because free-free cooling dominates. For a solar metallicity gas of density $n_\mathrm{H}=0.001 \ \mathrm{cm}^{-3}$, 
\ion{O}{VI} can be overabundant by an order of magnitude at $T  \approx 10^5 \ \mathrm{K}$. \ion{N}{V} behaves similarly as \ion{O}{VI}, but \ion{C}{IV} is underabundant relative to PIE at this density because of an overabundance in \ion{C}{V} at these cooler temperatures \citep[see also][for a comparison]{Oppenheimer:2013,Gnat:2017}.

\begin{figure*} 
  \centering
  \includegraphics[width=0.85\textwidth]{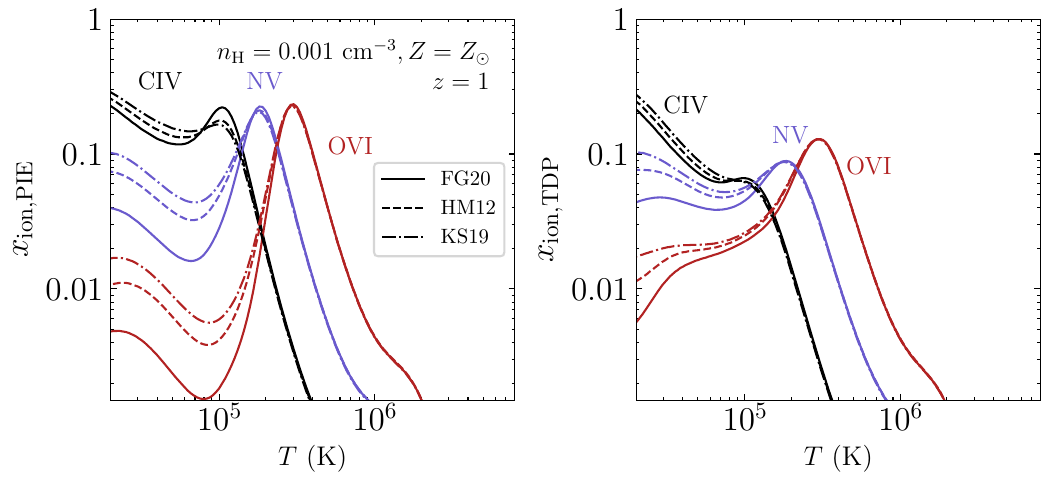}
  \caption{Impact of different UVBs in the predicted ion fractions under PIE and TDP conditions. The \textit{left} panel shows expected PIE ion fractions $x_{\rm ion}$ for \ion{C}{IV}, \ion{N}{V}, and \ion{O}{VI} assuming different UVB prescriptions. A fixed gas density of $n_\mathrm{H} = 0.001 \ \mathrm{cm}^{-3}$ and metallicity of $Z=Z_\odot$ are assumed for all calculations. The \textit{right} panel repeats the exercise from the left panel but for TDP conditions.}  \label{fig:TDP_UVB}
\end{figure*}

Since the publication of previous TDP calculations, the atomic data has been updated. Specifically, the cooling efficiencies have been updated across different versions of \texttt{CLOUDY}, which affects non-equilibrium fractions by extension because of the coupling between cooling losses and ionization state. Non-equilibrium calculations in Figure \ref{fig:TDP_PIE} were performed assuming the HM12 UVB and an initial temperature of $T_0 = 10^8 \ \mathrm{K}$ to enable direct comparisons with \cite{Gnat:2017} and \cite{Oppenheimer:2013}. The right panel of Figure \ref{fig:TDP_PIE} shows that all three studies produce consistent $x_{\rm ion}$ under TDP with differences of less than a factor of two at $T \lesssim 3 \times 10^4 \ \mathrm{K}$. This is an important caveat to consider when comparing the latest TDP calculations with those in the literature.

In addition to atomic data, the extragalactic ionizing background has also been updated since previous TDP calculations were performed (Figure \ref{fig:UVB_compare}). Therefore, there is a need to perform TDP calculations using updated KS19 and FG20 UVB prescriptions. The left panel of Figure \ref{fig:TDP_UVB} shows the expected PIE fractions under three different UVB prescriptions, which can differ by a factor of several at $T \lesssim 10^5 \ \mathrm{K}$ \citep[see also][]{Chen:2017, Zahedy:2019}.  The differences in the predicted $x_{\rm ion}$ of these intermediate- to high-ionization species can be attributed to the spectral slope of the adopted UVB with a softer spectrum (e.g., FG20) and lower intensities at high frequencies that led to lower high-ion fractions. The discrepancy due to different adopted UVB models continues to affect the predicted TDP fractions, as shown in the right panel of Figure \ref{fig:TDP_UVB}.

\subsection{Model assumptions} \label{sec:model_vary}

While TDP models provide a self-consistent framework for examining metal-enriched absorption systems, their underlying assumptions can have a notable effect on the expected ion fractions. Three key ingredients of time-dependent photoionization models are the assumed ionizing radiation, the initial temperature from which cooling occurs, and the adopted abundance pattern. Reasonable choices for each ingredient are explored in the subsections here.

\subsubsection{UVB prescription} \label{sec:TDP_UVB}

As discussed in \S\ \ref{sec:PIE}, there are several prescriptions available for the extragalactic UVB, which provides ionizing photons for the CGM. TDP models currently available in the public domain \citep[see e.g.,][]{Oppenheimer:2013,Gnat:2017} adopt UVB models from HM05 and HM12. New UVB prescriptions from KS19 and FG20 show better agreement with Ly$\alpha$ forest observations, among other improvements. However, TDP models using KS19 and FG20 are not publicly available.

The right panel in Figure \ref{fig:TDP_UVB} shows the predicted TDP fractions for \ion{C}{IV}, \ion{N}{V}, and \ion{O}{VI} for three different UVBs at $z=1$, including HM12, KS19, and FG20.  The predictions are obtained using the procedure described in \S\ \ref{sec:TDP}. The high-temperature ($T>10^5 \ \mathrm{K}$) peak comes from collisional ionization, while the low-temperature enhancement occurs because of recombination lags arising from rapid gas cooling. The fiducial choice of $n_\mathrm{H}=0.001 \ \mathrm{cm}^{-3}$ and $Z=Z_\odot$ for the density and metallicity are relevant to enriched \ion{C}{IV} absorbers. While ion fractions for \ion{C}{IV} are robust against UVB choices, \ion{N}{V} and \ion{O}{VI} TDP fractions can differ by a factor of up to three at $T \lesssim 10^5 \ \mathrm{K}$.  To facilitate a more focused discussion on the effect of TDP, FG20 is adopted as the fiducial UVB spectrum in the subsequent discussions.

\subsubsection{Initial temperature} \label{sec:TDP_T0}

A salient feature of TDP calculations is ion fractions' dependence on the gas's cooling history. This dependence is captured by solving the ionization and energy equations simultaneously. Therefore, the initial temperature from which the gas cools can affect TDP fractions at lower temperatures. Specifically, the gas is established to be in collisional ionization equilibrium (CIE) at the initial temperature $T_0$ at the beginning of TDP calculations (\S\ \ref{sec:TDP}). Assuming the gas is heated to $T_0 > 4 \times 10^5 \ \mathrm{K}$ by some initial process, the CIE assumption is justified because the collisional ionization time $t_\mathrm{ion} \sim 1/n_\mathrm{e} q_\mathrm{ion}$ \citep[where $q_\mathrm{ion}$ is the temperature-dependent collisional ionization coefficient, see e.g.][]{Voronov:1997} is less than the CIE cooling time $t_\mathrm{cool} \sim k_\mathrm{B}T/n_\mathrm{H} \Lambda$. At lower initial temperatures or with more efficient gas cooling \citep[due to the gas being underionized, see e.g.,][]{Gnat:2012}, CIE may not be established and subsequent non-equilibrium ion fractions will depend on the initial ionization state set by the relevant astrophysical process. Because the CIE composition is temperature-dependent \citep[see e.g.,][]{Gnat:2007}, different initial compositions corresponding to unique choices of initial temperature may influence the subsequent time-dependent cooling. Note that while dust grains can survive assuming $T_0 \lesssim 10^6 \ \mathrm{K}$ \citep[][]{Draine:2011}, fiducial \texttt{CLOUDY} simulations show that the inclusion of dust in non-equilibrium calculations does not affect ionic abundances for the species considered here \citep[see also][]{Montier:2004}.

\begin{figure} 
    \includegraphics[width=0.95\columnwidth]{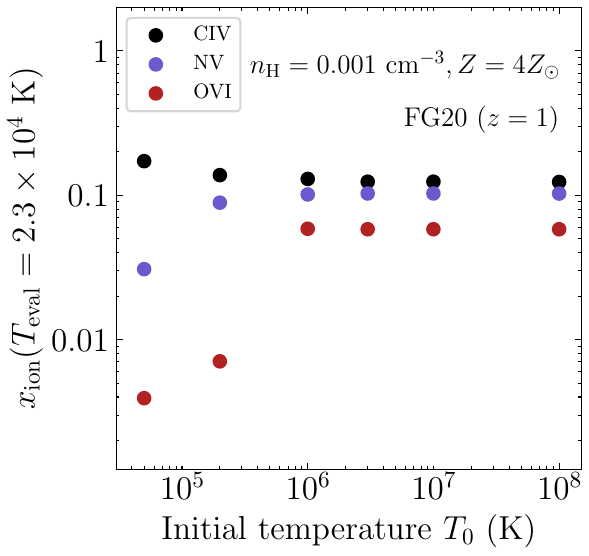}
  \caption{Dependence of TDP ion fractions $x_{\rm ion}$ on the initial gas temperature $T_0$ for the \ion{C}{IV}, \ion{N}{V}, and \ion{O}{VI} ions.  The resulting $x_{\rm ion}$ are evaluated at $T_\mathrm{eval} = 2.3 \times 10^4 \ \mathrm{K}$. The FG20 UVB at $z=1$ is adopted, and the density and metallicity are assumed to be $n_\mathrm{H} = 0.001 \ \mathrm{cm}^{-3}$ and $Z=4Z_\odot$.} \label{fig:TDP_T0}
\end{figure}

\begin{figure*} 
  \centering
  \includegraphics[width=0.85\textwidth]{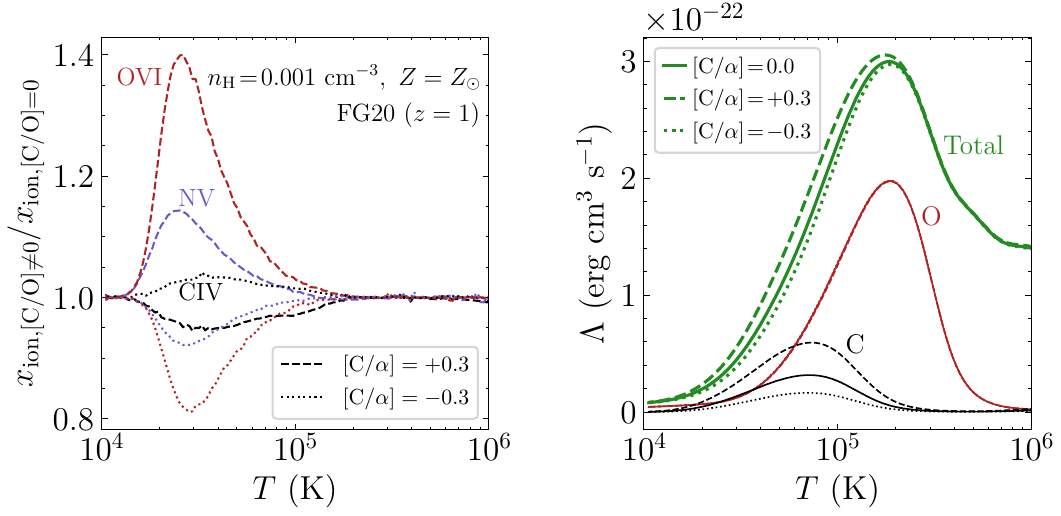}
  \caption{Impact of non-solar $\mathrm{[C/\alpha]}$ relative abundances on the predicted ion ratios. The \textit{left} panel depicts the ratio of TDP fractions evaluated assuming non-solar $\mathrm{[C/O]}$ ratios relative to the standard solar values. Examples with super- and sub-solar $\mathrm{[C/\alpha]}$ are shown with different line styles. The FG20 UVB at $z=1$ is adopted, along with a density of $n_\mathrm{H} = 0.001 \ \mathrm{cm}^{-3}$ and metallicity of $Z=Z_\odot$. The \textit{right} panel depicts TDP cooling losses assuming a non-solar $\mathrm{[C/\alpha]}$ ratio. Individual contributions from carbon and oxygen are depicted in different colors. The total includes contribution from nitrogen, $\alpha$-elements like neon and sulfur, as well as light elements like hydrogen and helium.} \label{fig:TDP_abund}
\end{figure*} 

Figure \ref{fig:TDP_T0} depicts the TDP fractions for key species evaluated at $T_\mathrm{eval} = 2.3 \times 10^4 \ \mathrm{K}$ (the line width temperature of the absorber discussed in \S\ \ref{sec:analysis}) for different choices of initial temperature $T_0$. For reference, \cite{Gnat:2017} assume $T_0 = 10^8 \ \mathrm{K}$ for their TDP calculations, while \cite{Oppenheimer:2013} experiment with two temperatures, $T_0 = 3 \times 10^6, 10^7 \ \mathrm{K}$. It is found that the evaluated $x_{\rm ion}$'s are robust for initial temperatures of $T_0 \ge 10^6 \ \mathrm{K}$. However, the \ion{O}{VI} fraction is depleted for $T_0 \lesssim 10^6 \ \mathrm{K}$ by an order of magnitude, and the anticipated \ion{N}{V} fraction decreases for $T_0 \lesssim 10^5 \ \mathrm{K}$ by a factor of three, while the \ion{C}{IV} fraction remains nearly stable for the range of initial temperatures considered. The sensitivities of TDP ion fractions to different initial temperature can be understood by considering the CIE fractions of different ionization stages at these temperatures for different elements.  For oxygen at initial temperatures cooler than $T=2\times 10^5$ K, a substantial fraction remains at lower ionization stages under CIE, resulting in a lower initial O$^{5+}$ abundance at the start of the cooling sequence and a still lower $x_{\rm ion}$ as the gas continues to cool. Note that the effect of the adopted initial temperature is metallicity-dependent. For instance, adopting a solar metallicity leads to a 40\% decrease in the \ion{O}{VI} fraction for $T_0 \lesssim 10^6 \ \mathrm{K}$ as opposed to depletion by an order of magnitude for $Z=4Z_\odot$ (the 1-$\sigma$ upper bound on the metallicity of the absorber discussed in \S\ \ref{sec:analysis}) shown in Figure \ref{fig:TDP_T0}.

The physical justification for exploring different initial temperatures is related to the possible dependence of the initial temperature with halo mass. A potential scenario is infalling clouds getting shock-heated to the halo's virial temperature, giving rise to these intermediate- to high-ionization species. Galaxy hosts of these absorbers having a large range of halo masses \citep[see e.g.,][]{Liang:2014, Bordoloi:2014, Burchett:2013, Zahedy:2019, Qu:2023}, indicating a range of initial temperatures from which absorbers cool down under the infall scenario. Cooling outflows may also affect the initial temperature, especially for dwarf galaxies where coronal \ion{O}{VI} observations have been reported despite lower expected virial temperatures \citep[see e.g.,][]{Heckman:2001}. Allowing flexibility in choosing the initial temperature for TDP calculations can thus help incorporate known galaxy host information in modeling these absorbers. 

\subsubsection{Elemental abundance pattern} \label{sec:TDP_abund}

Chemical elements are produced through different nucleosynthetic pathways. The $\alpha$ elements (O, S, Mg, Si, Ne, etc.) primarily originate in core-collapse supernovae. However, carbon can have secondary contributions from AGB winds \citep[see e.g.,][]{Kobayashi:2020}. For these reasons, both the $\alpha$ elements and carbon can be expected to deviate from the solar expectation. The relative abundances, including $\mathrm{[C/\alpha]}$, $\mathrm{[N/\alpha]}$, and $\mathrm{[Fe/\alpha]}$, encapsulate the deviation of different elements from the solar pattern. The relative abundances of elements set their contributions to gas cooling \citep[see e.g.,][]{Gnat:2012}. Because $x_{\rm ion}$'s in PIE are evaluated independently of the gas cooling under a steady state assumption, relative abundances do not influence PIE fractions. However, ion fractions are coupled to gas cooling in TDP, meaning that relative abundances can affect TDP fractions.

Figure \ref{fig:TDP_abund} shows the variation in TDP fractions and gas cooling using non-solar $\mathrm{[C/\alpha]}$ values. Note that for this exercise, all elements except carbon were assumed to be at the solar level, and carbon was scaled up or down with respect to the rest of the elements. Despite carbon being an effective coolant, variations in $\mathrm{[C/\alpha]}$ have a small impact on the total gas cooling and TDP fractions. This is because at $T \approx 4 \times 10^7 \ \mathrm{K}$, which is the cooling peak of carbon, it only contributes to about 20\% of the total cooling (assuming a solar pattern).
The total cooling for $\mathrm{[C/\alpha]} = \pm 0.3$ is therefore respectively 20\% higher and 10\% lower compared to the original. The coupled $x_\mathrm{ion}$'s change by up to 40\%. Nevertheless, the relative abundances must be accounted for when computing TDP column densities before comparing them with absorption line measurements between the ionization states of different elements.

\section{Applications to empirical data} \label{sec:analysis}

As noted in \S\ \ref{sec:ion_models}, a \ion{C}{IV} absorber at $z_\mathrm{abs} \approx 1.26$ was found by \cite{Kumar:2024} to exhibit super solar metallicity with an anticipated gas temperature significantly cooler than what is inferred from the thermal width of the absorption lines. This system consists of a single narrow component of $\log\,N_c({\rm CIV})/\cmjj=12.78\pm 0.03$ and $b_c({\rm CIV}) = 6.0\pm 0.7 \ \kms$ in the high signal-to-noise Keck HIRES spectrum. Associated absorption features in \ion{C}{III} and \ion{O}{IV} were identified in the \textit{HST} STIS and COS spectrum, respectively, yielding a density estimate $\log(n_\mathrm{H}/\mathrm{cm}^{-3}) \approx -3.3\pm0.2$ and relative abundance $\mathrm{[C/\alpha]} \approx 0.0$ under a PIE assumption. Including the best estimated \ion{H}{I} column density of $\log\,N_c({\rm HI})/\cmjj=12.9\pm 0.1$, the observed relative ion to hydrogen ratios lead to a high inferred metallicity of $\mathrm{[\alpha/H]}=+0.6_{-0.1}^{+0.2}$ under PIE, approximately $4\times$ solar with a solar metallicity ruled out at a 6-$\sigma$ level of significance. No luminous galaxies are found in the vicinity of the absorber and the presence of faint galaxies with unobscured SFR of $\lesssim 1 M_\odot\mathrm{yr}^{-1}$ was ruled out at the 3-$\sigma$ limit in \cite{Kumar:2024} using the QSO light subtracted MUSE cube.

\begin{table} 
\begin{threeparttable}
\caption{Inferred physical properties for component c3 of the $z_\mathrm{abs} = 1.25937$ absorber presented in \cite{Kumar:2024}.} \label{tab:z_126}
\setlength\tabcolsep{0pt} 
\footnotesize\centering
\smallskip 
\begin{tabular*}{\columnwidth}{@{\extracolsep{\fill}}ccc}
\toprule
Property & TDP \tnote{a} & PIE \\
\midrule
\midrule
$\log(n_\mathrm{H}/\mathrm{cm}^{-3})$ & $-3.2 \pm 0.1$ & $-3.3 \pm 0.2$ \\ 
$T \ (\mathrm{K})$ & $2.3 \times 10^4$ \tnote{a} & $<1.3 \times 10^4$ \tnote{b} \\ 
$\mathrm{[\alpha/H]}$	& $0.4 \pm 0.2$ &	$0.6_{-0.1}^{+0.2}$ \\
$\mathrm{[C/\alpha]}$	& $0.25 \pm 0.08$ & $0.00 \pm 0.09$ \\
$\mathrm{[N/\alpha]}$	& $<0.8$ & $<0.6$ \\
$\log(l/\mathrm{kpc})$	& $-1.8_{-0.2}^{+0.3}$ & $-1.9 \pm 0.3$ \\
\bottomrule
\end{tabular*}
\begin{tablenotes}\footnotesize
\item[a] Line-width determined temperature.
\item[b] 3-$\sigma$ upper limit on the PIE temperature. The best-fit PIE models suggest that the gas cools to a median temperature of $\approx 5000$ K
\end{tablenotes}
\end{threeparttable}
\end{table}

The posterior distribution for the gas density and metallicity were used to obtain the predicted ionization model temperatures, $T_\mathrm{PIE}$. Owing to the high gas metallicity, ionization models prefer a very cool gas with median equilibrium temperature $T_\mathrm{PIE} \approx 5000 \ \mathrm{K}$, and a stringent 3-$\sigma$ upper limit of $13000 \ \mathrm{K}$ (see Table \ref{tab:z_126}). The ionization model temperature is inconsistent with the thermal temperature $T \approx 23000 \ \mathrm{K}$ obtained by comparing the \ion{H}{I} line width, $b_c({\rm HI}) = 20\pm 2\,\kms$, with the metal absorption. The coincidence of this discrepancy with the high inferred metallicity of this component motivates its reassessment under the TDP framework discussed below.

\begin{figure} 
  \centering
  \includegraphics[width=\columnwidth]{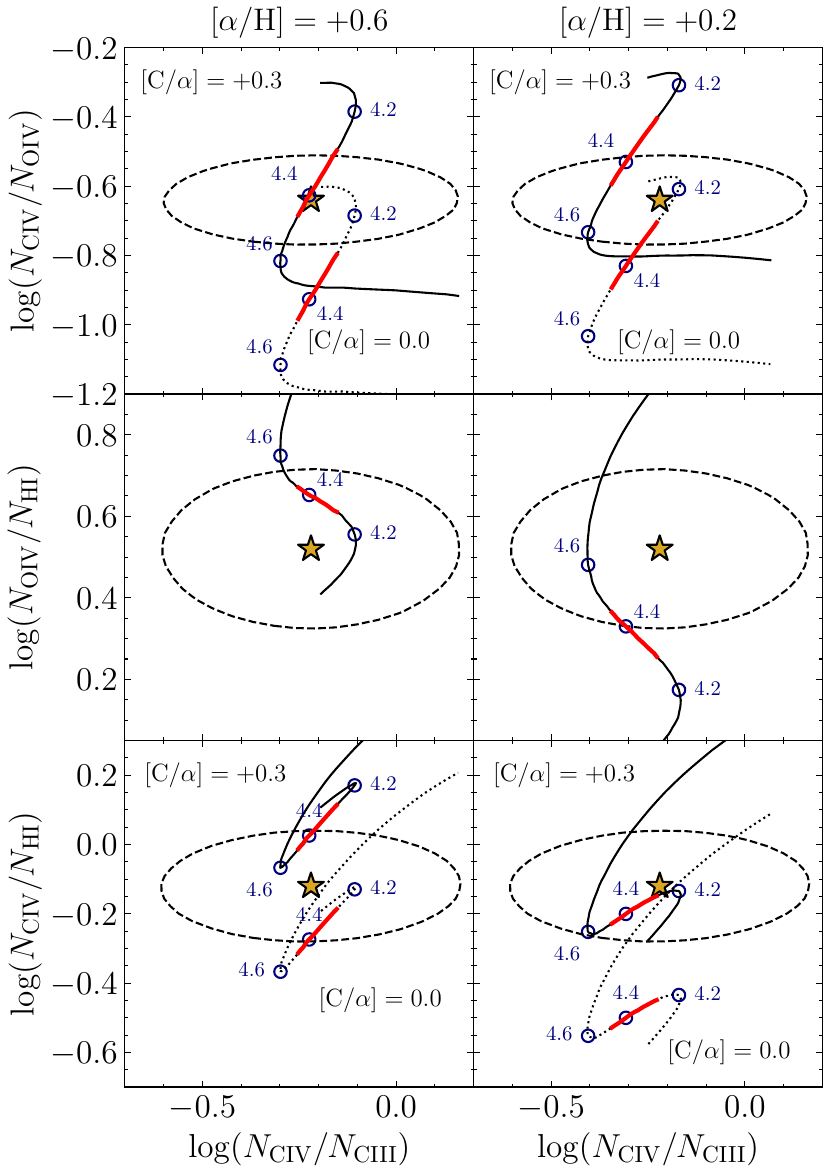}
  \caption{Re-evaluation of a \ion{C}{IV} absorber at $z\approx 1.26$ from \cite{Kumar:2024} under TDP and non-solar abundance pattern. From top to bottom, the panels show respectively observed and predicted \ion{C}{IV}/\ion{O}{IV}, \ion{O}{IV}/\ion{H}{I}, and \ion{C}{IV}/\ion{H}{I} versus the \ion{C}{IV}/\ion{C}{III} ratio. The observed values are marked by a star symbol, with the corresponding 68\% error ellipse drawn with a dashed line. TDP model predictions in each panel are shown for a fiducial gas density of $\log(n_\mathrm{H}/\mathrm{cm}^{-3}) = -3.2$ and metallicities of $\mathrm{[\alpha/H]}=+0.6,+0.2$ (left and right columns respectively) at different temperatures with the corresponding $\log\,T/{\rm K}$ marked in blue color.  The 1-$\sigma$ interval in temperature inferred from line width measurements is highlighted in thick red segments. For ratios \ion{C}{IV}/\ion{O}{IV} and \ion{C}{IV}/\ion{H}{I}, the model predictions also depend on the relative carbon abundance.  Therefore, two sets of model predictions based on $\mathrm{[C/\alpha]} = 0.0$ and $+0.3$ are shown in dotted and solid curves, respectively.
}  \label{fig:z_126}
\end{figure}
 
Unlike PIE, thermal equilibrium is not assumed in TDP. A comparison between predicted and observed ionic column densities of this component is performed for a gas temperature of $T_\mathrm{TDP}=23000\,{\rm K}$ as determined from comparing absorption line widths of different elements. Note that the TDP model grid used here is built assuming that the gas cools from an initial temperature of $T_0 = 3 \times 10^6 \ \mathrm{K}$. The model predictions are calculated over a grid of gas parameters. The conversion from TDP fractions $x_\mathrm{ion}$ to column densities is achieved by assuming $\log(N_\mathrm{HI}/\mathrm{cm}^{-2}) = 12.9$ \citep[see Equation (3) from][]{Kumar:2024}.  A posterior distribution is constructed for gas parameters using MCMC sampling utilities provided by the \texttt{emcee} library in \texttt{Python} \citep[][]{F-M:2013}. The likelihood definition used for this evaluation is taken from \cite{Kumar:2024}, wherein sensitive upper limits on column density for non-detections are accounted for in addition to measurements. The best-fit values and associated uncertainties for various gas properties are summarized in Table \ref{tab:z_126}. Estimates for both the TDP and PIE frameworks are listed in the table to facilitate direct comparisons. 

The best-fit model parameters summarized in Table \ref{tab:z_126} confirm that relaxing the equilibrium condition reduces the inferred gas metallicity. Although the best-fit gas metallicity remains high at [$\alpha$/H]$=+0.4$, solar metallicity cannot be ruled out at a significant level.  In addition, the requirement for a non-solar C/$\alpha$ elemental abundance pattern of [C/$\alpha$]$=0.25$ is robustly confirmed based on the observed abundances of carbon ions.

To further visualize the allowed parameter range, Figure \ref{fig:z_126} illustrates how the observed ionic column density ratios depend on the gas temperature. The $x$-axis in all panels is chosen to be the \ion{C}{IV}/\ion{C}{III} ratio, which has the benefit of being independent of the underlying abundance pattern despite a large uncertainty in $N({\rm CIII})$.  The observed column density ratios and associated uncertainties are respectively marked by a star symbol and a dashed ellipse in each panel.  The model curves track the changes in the ionization state of the gas as it cools. Model predictions are shown for the best-fit gas density of $\log(n_\mathrm{H}/\mathrm{cm}^{-3})=-3.2$ and the 1-$\sigma$ upper- and lower-bound of the best-fit metallicity, $\mathrm{[\alpha/H]} = +0.6,+0.2$. Column density ratios for the allowed temperature range (68\% interval around the best-fit line-profile temperature) are indicated in red. The two bottom panels in the left column of Figure \ref{fig:z_126} demonstrate that with a solar abundance pattern (dotted curve), the best-fit TDP model with significantly enhanced gas metallicity $\mathrm{[\alpha/H]} = +0.6$ can already reproduce the observed \ion{O}{IV}/\ion{H}{I} and \ion{C}{IV}/\ion{H}{I} ratios to within the measurement uncertainties, while it under-produces the observed \ion{C}{IV}/\ion{O}{IV} ratio displayed in the top panel. By contrast, a TDP model with only slightly enhanced gas metallicity of $\mathrm{[\alpha/H]} = +0.2$ assuming a solar abundance pattern (right column of Figure \ref{fig:z_126}) can reproduce the observed \ion{C}{IV}/\ion{O}{IV} and \ion{O}{IV}/\ion{H}{I} ratios, but under-produces the \ion{C}{IV}/\ion{H}{I} ratio. To mitigate the discrepancy between observed and model \ion{C}{IV}/\ion{O}{IV} and \ion{C}{IV}/\ion{H}{I} ratios (at $\mathrm{[\alpha/H]}=+0.6,+0.2$ respectively) would require increasing the relative carbon abundance to [C/$\alpha$]$\approx+0.3$ (solid curve in all panels).  A still higher [C/$\alpha$] value would increase the tension between the observed and predicted \ion{C}{IV}/\ion{H}{I} and \ion{C}{IV}/\ion{O}{IV} ratios displayed in the left and right columns respectively.

\section{Summary and Discussion} \label{sec:discussion}

This paper presents an updated framework for computing time-dependent photoionization models, integrating the latest UVB prescriptions and atomic data.  Specifically, the study focuses on high-metallicity gas, in which rapid cooling results in a recombination lag and enhances the abundances of higher ionization states. The findings demonstrate that accounting for the possibility of highly enriched CGM absorbers being out of equilibrium helps resolve the discrepancy between the thermal temperatures derived from absorption line profiles and the photoionization temperatures predicted by best-fit equilibrium ionization models. In addition, it minimizes the necessity of invoking uncharacteristically high super-solar gas metallicities to explain the observed ion abundances.  Under TDP, solar metallicity cannot be ruled out with more than 2-$\sigma$ significance, while a non-solar $\mathrm{[C/\alpha]} = 0.25 \pm 0.08$ is robustly confirmed using the observed relative ion abundances.  Extensive studies have also investigated the impact of different model assumptions involving the choice of a particular UVB prescription, the initial temperature, and the adopted elemental abundance pattern. The lessons learned from this exercise are summarized here.  

First, the expected abundances of common ionization species traced by \ion{C}{IV}, \ion{N}{V}, and \ion{O}{VI} are sensitive to the slope and amplitude of the adopted UVB (see Figure \ref{fig:TDP_UVB} and \S\ \ref{sec:TDP_UVB}).  This sensitivity arises because higher radiation intensities result in increased abundances of ions with ionization potentials matching the photon energies at these high frequencies (see e.g., Figure \ref{fig:UVB_compare}). Notably, this effect persists regardless of whether the gas is in equilibrium at these densities.

Secondly, the initial temperature $T_0$ is expected to influence the subsequent production of intermediate- to low-ionization species under a rapidly cooling process. $\mathrm{O}^{5+}$ ions exhibit appreciable change (by an order of magnitude) in the predicted TDP fractions when $T_0=10^5$--$10^6$ K was adopted, assuming that the gas follows CIE at $T_0$. Meanwhile, N$^{4+}$ ions deplete by a factor of three, and no detectable differences are found for C$^{3+}$ with $T_0>10^5$ K (see Figure \ref{fig:TDP_T0} and \S\ \ref{sec:TDP_T0}).  The sensitivity of the predicted $\mathrm{O}^{5+}$ abundances under TDP can be explained by the expectation that under equilibrium conditions the abundances of these ions peak at a temperature higher than $T_0$ and, therefore, a lower initial abundance is expected at $T_0$. Conversely, C$^{3+}$ or N$^{4+}$ have peak temperatures comparable to or lower than the initial temperatures considered, rendering their resulting TDP fractions insensitive to the adopted $T_0$. This effect is also metallicity-dependent -- the sensitivity of TDP fractions to the adopted initial temperature reduces for lower gas metallicities.

Finally, as illustrated in Figure \ref{fig:TDP_abund}, cooling at $T>10^5$ K is driven by metal ions. The expected TDP ion fractions should naturally depend on the underlying elemental abundance pattern. Indeed, changing the relative carbon to $\alpha$-element abundance pattern can lead to different expected ion fractions by 5-40\% at $T<10^5$ K under the TDP scenario, with the largest anticipated differences for \ion{O}{VI} (\S\ \ref{sec:TDP_abund}).  The relatively modest effect under a non-solar [C/$\alpha$] pattern arises because carbon only contributes to $\approx 20 \%$ of the total cooling at its peak, $T \approx 4 \times 10^7 \ \mathrm{K}$ assuming a solar pattern.

However, caveats remain.  A direct consequence of non-equilibrium conditions is that the gas is in a transient state as it continues to cool, posing challenges in constructing a steady-state model for characterizing the origin of these metal-line absorbers as a whole. Alternatively, the temperature discrepancy under PIE can be addressed by introducing an unknown heating source to elevate the heating rate (see Figure \ref{fig:heat_cool}) and maintain the temperature at the value suggested by thermal broadening during photoionization calculations (e.g., \citealt{Hussain:2015, Pachat:2017}; see also \citealt{Qu:2023}). The challenge in this scenario is maintaining the required heating level without significantly modifying the relative ion abundances.  While turbulence may be a suitable heating agent as it does not generate ionizing photons, the small non-thermal broadening observed in the case study (\S\ \ref{sec:analysis}) with $b_\mathrm{NT} < 6 \ \kms$ (a 3-$\sigma$ limit, see Appendix A5 in \cite{Kumar:2024}) does not support this scenario.

Irrespective of the challenges associated with TDP and the warm PIE models, current observations do not provide the precision needed to discriminate between the two scenarios. Because recombination lags are expected to result in an overabundance of higher ionization species from non-equilibrium cooling, producing characteristically different abundances of successive ionization stages compared to the equilibrium scenario, improved measurements of multiple ionization stages with reduced uncertainties help strengthen the distinguishing power between PIE and TDP solutions statistically. Extending the TDP analyses to a larger sample of super-solar absorption components with high signal-to-noise data for successive ionization stages is needed to better quantify the fraction of metal absorbers originating in rapid cooling gas.

\section*{Acknowledgements}

The authors thank an anonymous referee for constructive comments that helped improve the presentation of this work. The authors also thank Zhijie Qu, Harley Katz, and Gwen Rudie for helpful discussions throughout this work. S.K.\ thanks Mu-Chen Hsieh and Rohan Venkat for testing the user-friendliness of open source \texttt{CLOUDY} scripts and \texttt{Python} routines used to create model grids used in this work. H.-W.C. and S.K.\ acknowledge partial support from HST-GO-17517.01A and NASA ADAP 80NSSC23K0479 grants.

\bibliography{TDP}
\bibliographystyle{mnras}

\clearpage
\clearpage

\end{document}